\documentclass[
    ,final            
  ]
  {aipproc}

\usepackage{epsfig}

\layoutstyle{6x9}

\begin{document}

\title{Transverse Spin Results From PHENIX}


\author{Feng Wei for the PHENIX collaboration}{
  address={New Mexico State University, Las Cruces, New Mexico, 88003, USA}
}

\begin{abstract}
The PHENIX experiment at the Relativistic Heavy Ion Collider explores the spin
structure of the proton in polarized p+p collisions at center-of-mass energies up to 500 GeV.
Tremendous experimental and theoretical progress has been made toward understanding the
physics involved with transversely polarized beams or targets in recent years. Not only nucleon
structure and parton distribution functions but also QCD dynamics have been studied in various
physics processes in high-energy polarized DIS and p+p collisions. In the 2006 and 2008 RHIC
runs, the PHENIX experiment took a significant amount of transversely polarized p+p collision
data at 200 GeV center-of-mass energy, with an integrated luminosity of 8 $pb^{-1}$ and beam
polarizations up to 50\%. Single spin asymmetries of different probes have been measured in
mid- and forward-rapidities. In this report, we present the latest transverse spin results from the
PHENIX experiment and discuss briefly the prospects of future transverse spin physics with the
PHENIX detectors upgrades.

\end{abstract}

\maketitle


\section{INTRODUCTION}
Transverse spin physics has attracted huge interests after Fermilab E704 experiment
observed large transverse single spin asymmetry (TSSA) in high-energy p+p collisions
which was believed to be very small based on pQCD arguments. So far, it has been
learned that two effects known as the Sivers \cite{Sivers} and Collins \cite{Collins} effects may have
contribution to the large TSSAs. However, the QCD mechanisms for these effects are
very complicated and still not quite clear in high energy p+p collision \cite{Kang1}.

As the first and only polarized p+p collider in the world, the Relativistic Heavy Ion
Collider (RHIC) has opened up a new energy regime in which to study the spin
structure of the proton. The PHENIX experiment, one of the two ongoing experiments
at RHIC, has been taking data with transversely polarized proton-proton collisions
since 2002. In the running of 2006 and 2008, PHENIX had accumulated a large data
sample with integrated luminosity of 8 $pb^{-1}$ and beam polarization up to 50\%, and a
number of new measurements have become possible with increased statistics.
\section{PHENIX EXPERIMENTAL SETUP}
The PHENIX detector \cite{detector} consists of central and forward spectrometer arms.
The central detector at mid-rapidity ($|\eta| < 0.35$) has two arms covering $2 \times
\frac{\pi}{2}$ in azimuth and is able to detect photons, neutral pions, identified charged hadrons,
electrons and $J/\psi$ mesons. The forward detector consists of muon spectrometers and
muon piston calorimeters (MPC). The muon spectrometers have a full $2\pi$ coverage at
$1.2<|\eta|< 2.4$ and can track muons and unidentified charged hadrons. The MPC is able
to measure photons and neutrol pions within $3.1 < |\eta| < 3.9$.

\section{PHENIX RESULTS}
\subsection{$A_N$ for light hadrons at mid-rapidity and forward rapidity}
\begin{figure}
  \psfig{file=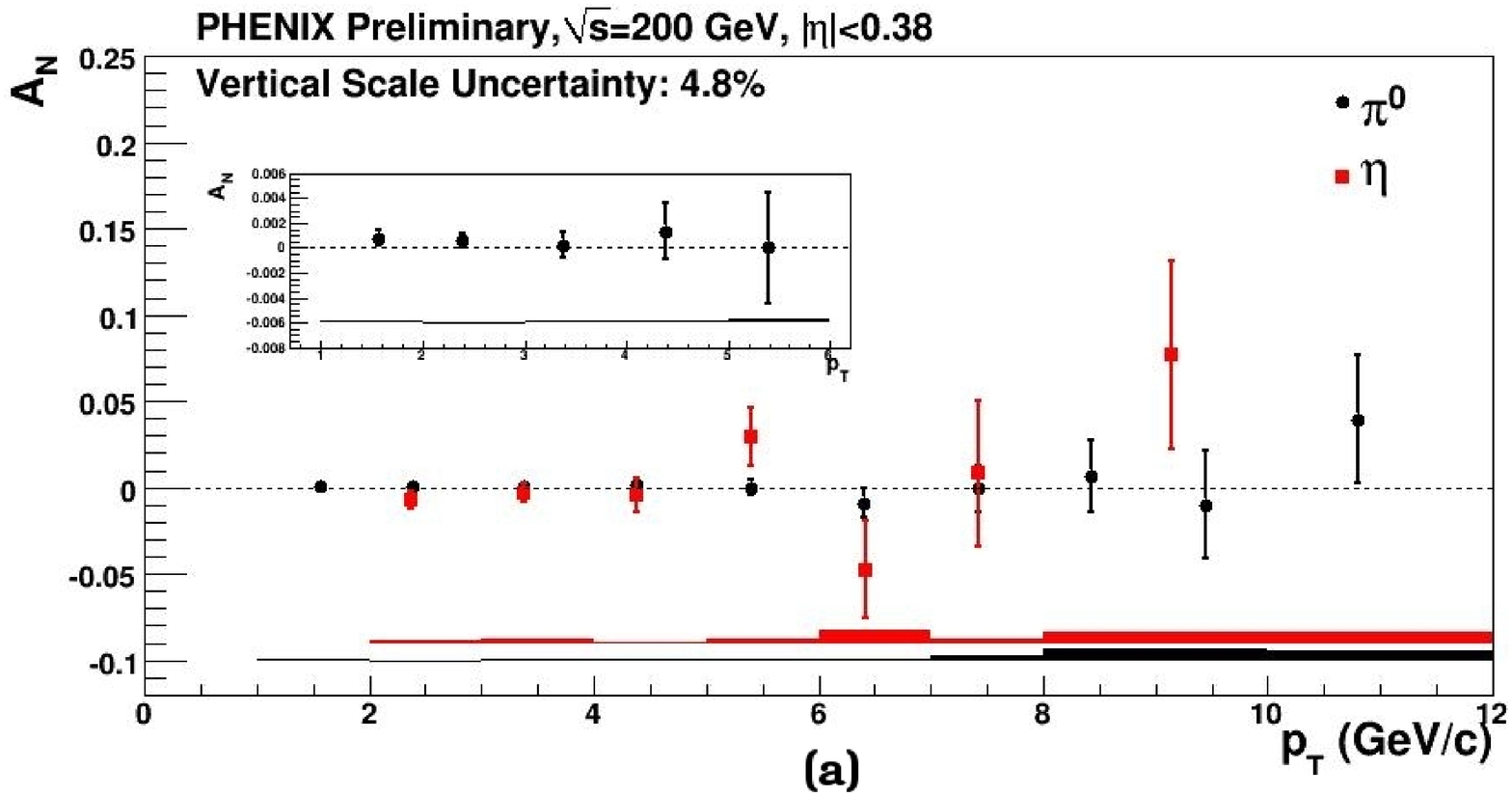, width=0.5\textwidth}
  \psfig{file=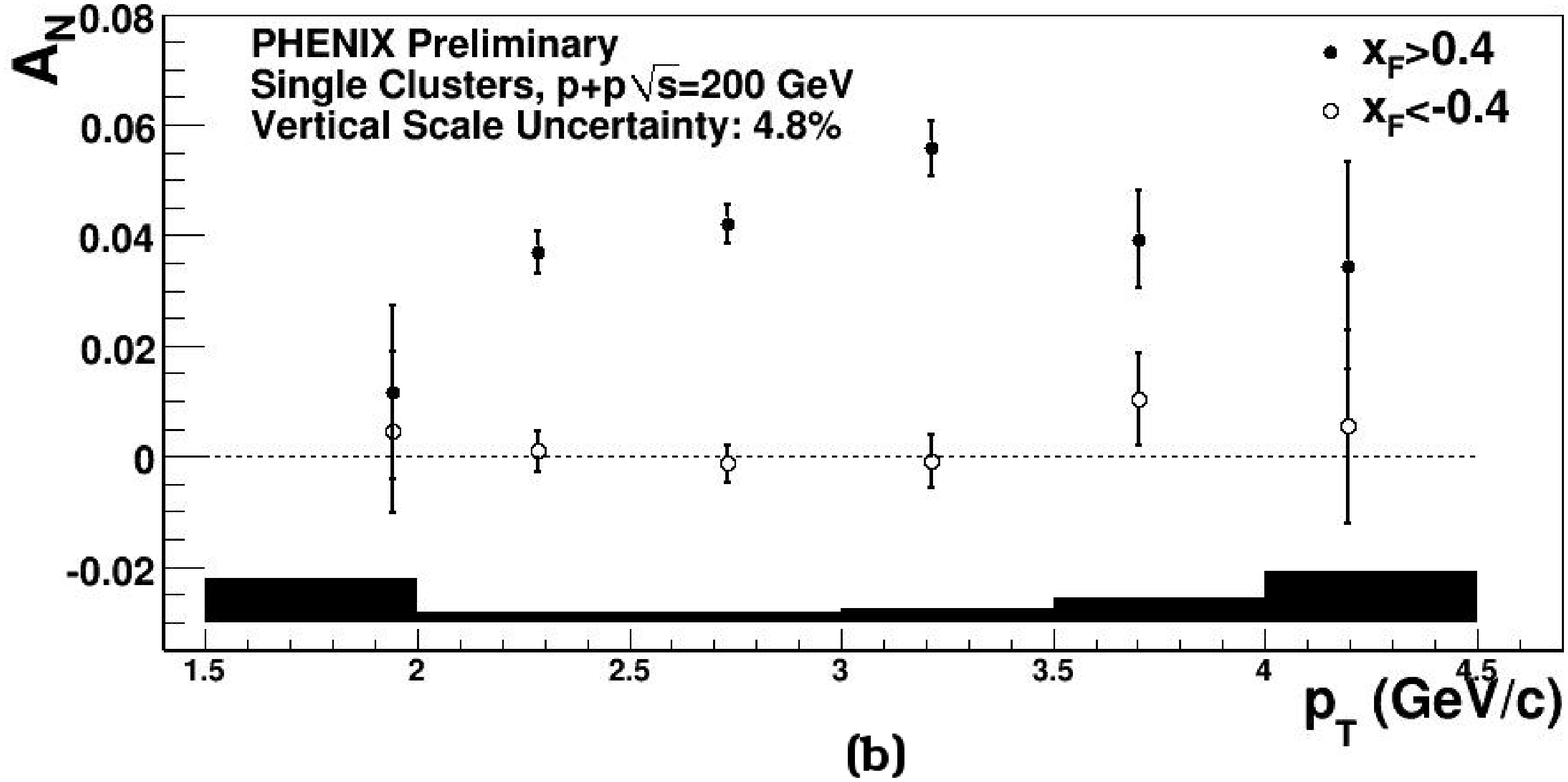, width=0.5\textwidth}
  \caption{(a) $A_N$ of $\pi^0$ and $\eta$ production at mid-rapidity. (b) $A_N$ vs $p_T$ for single clusters measured by the MPC.}
\label{fig_Run8_pi0}
\end{figure}

Low transverse momentum ($p_{T}$) hadron production at mid-rapidity is dominated by
gluon-gluon and quark-gluon processes, which can be used to constrain the gluon
Sivers effect in TSSA measurements. The first TSSA result in $\pi^0$ and $h^{+/-}$ production at
mid-rapidity has been published by using the 2002 data sample \cite{mid2002}. Figure \ref{fig_Run8_pi0}-(a) shows the
$\pi^0$ and $\eta$ TSSAs measured in 2008, which have a figure of merit 300 times higher than
the one in 2002.

The new MPC detectors installed in 2006 measure inclusive photons in the very
forward rapidity. Very significant non-zero TSSAs have been measured for single
clusters that mix neutral pions and inclusive photons by the MPC in 2008. Especially
in this measurement, the turning of $p_T$ dependent TSSAs was observed as shown in
Figure \ref{fig_Run8_pi0}-(b), which could become an important verification to the $1/p_T$ prediction made by
the twist-3 mechanism if better statistics would be achieved in future.

\begin{figure}
  \psfig{file=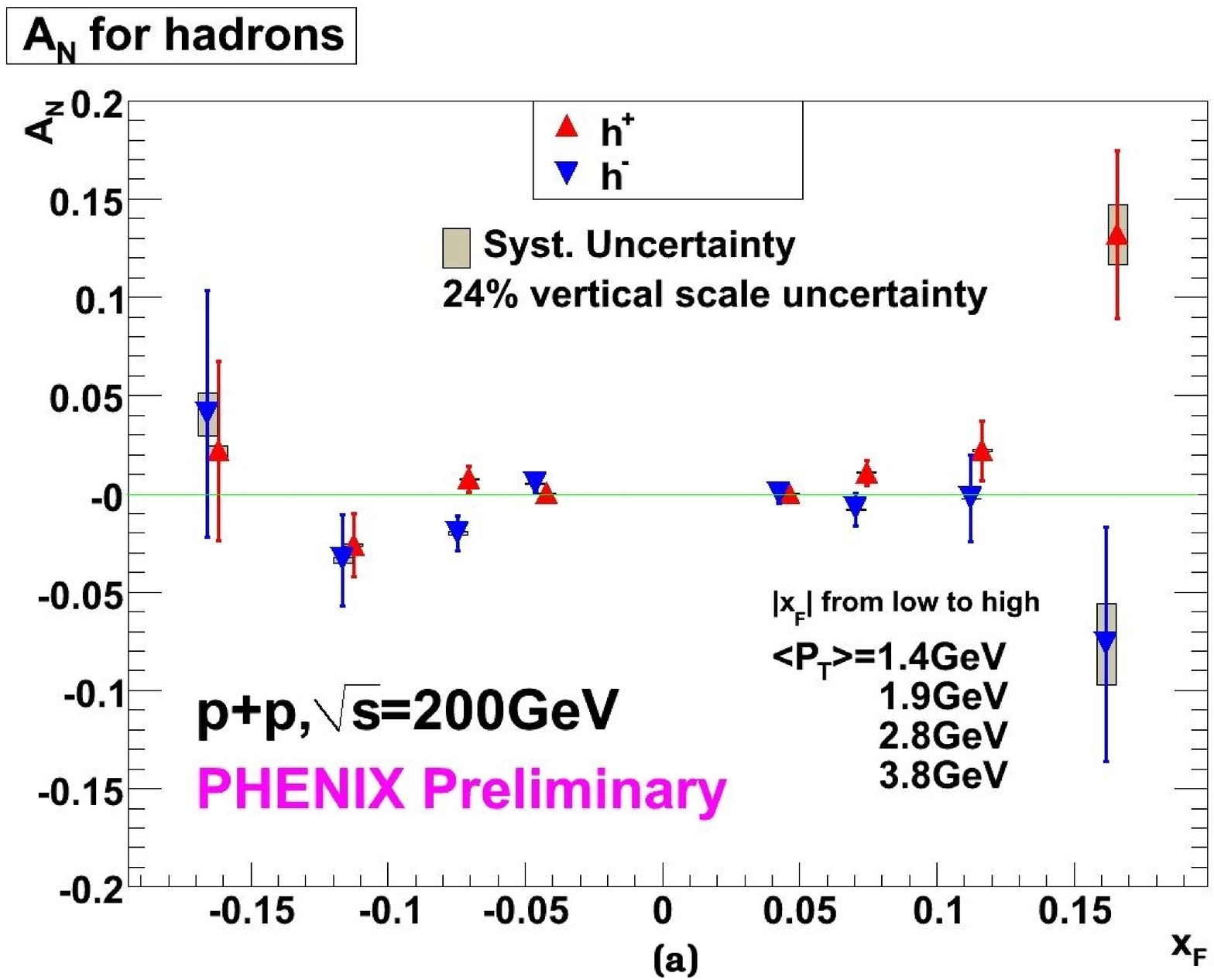, width=0.5\textwidth}
  \psfig{file=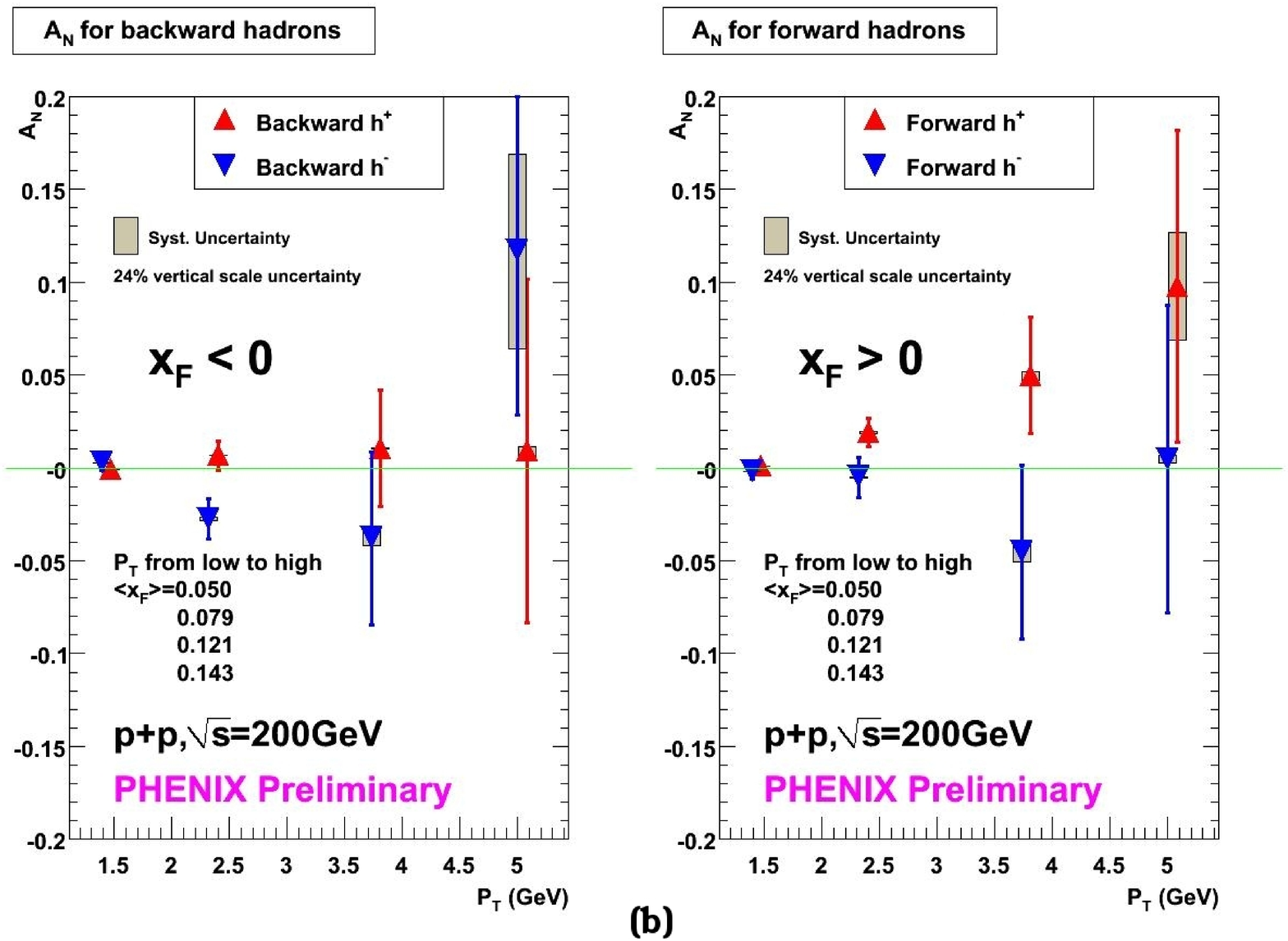,width=0.5\textwidth}
  \caption{$A_N$ of charged hadrons measured by muon spectrometers. (a) $A_N$ vs $x_F$, (b) $A_N$ vs $p_T$.}
\label{fig_Run8_SingleHadron}
\end{figure}

The muon spectrometers are capable of measuring inclusive charged hadrons at
forward rapidity where large TSSAs were observed by other experiments when new
hadron trigger has been implemented in 2008. Non-zero TSSAs have been measured
in unidentified charged hadron production at relatively lower $x_F$ (the longitudinal
fractional particle momentum) and higher $p_T$ region, which confirmed previous results
from BRAHMS, see Figure \ref{fig_Run8_SingleHadron}.
\subsection{$A_N$ for heavy flavor at forward rapidity}
\begin{figure}
  \psfig{file=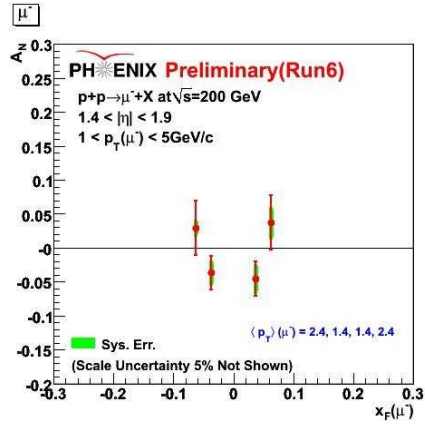,width=0.4\textwidth}
  \caption{$A_N$ for single muons from heavy flavor decays measured by the muon spectrometers.}
\label{fig_Run6_Single_Muon_AN}
\end{figure}

Besides the traditional light hadron probes, PHENIX also carried out the
measurements of TSSAs in heavy flavor quark production through decay muons by
using 2006 dataset as Figure \ref{fig_Run6_Single_Muon_AN} shows. This channel is sensitive to the gluon Sivers
effect because gluon-gluon fusion is the dominant process at RHIC energy for $D$
meson production. The result of TSSA in heavy flavor production that is dominant by
$D$ meson in the PHENIX experiment is consistent with zero.
\subsection{Di-hadron correlation at mid-rapidity}
\begin{figure}
  \psfig{file=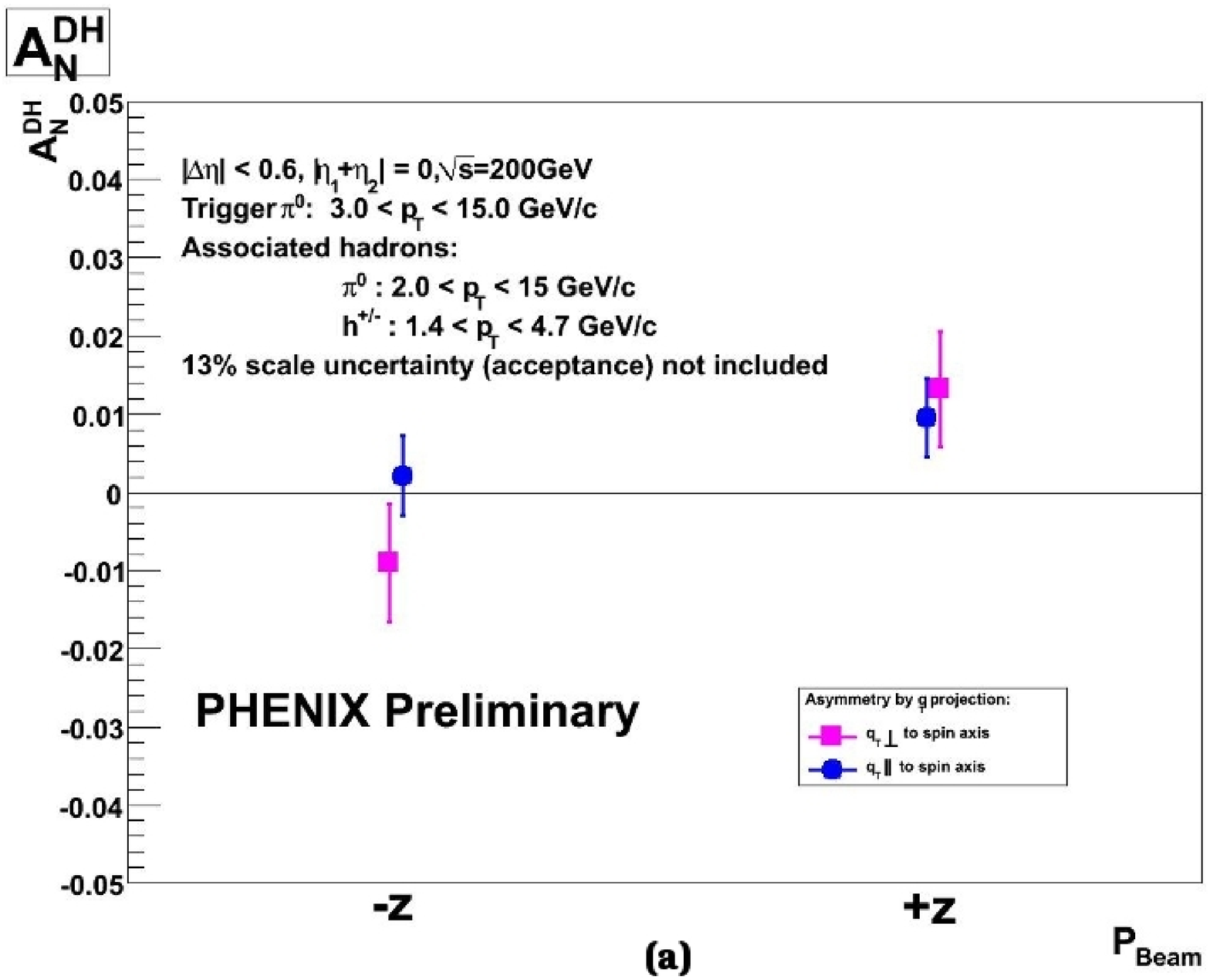, width=0.5\textwidth}
  \psfig{file=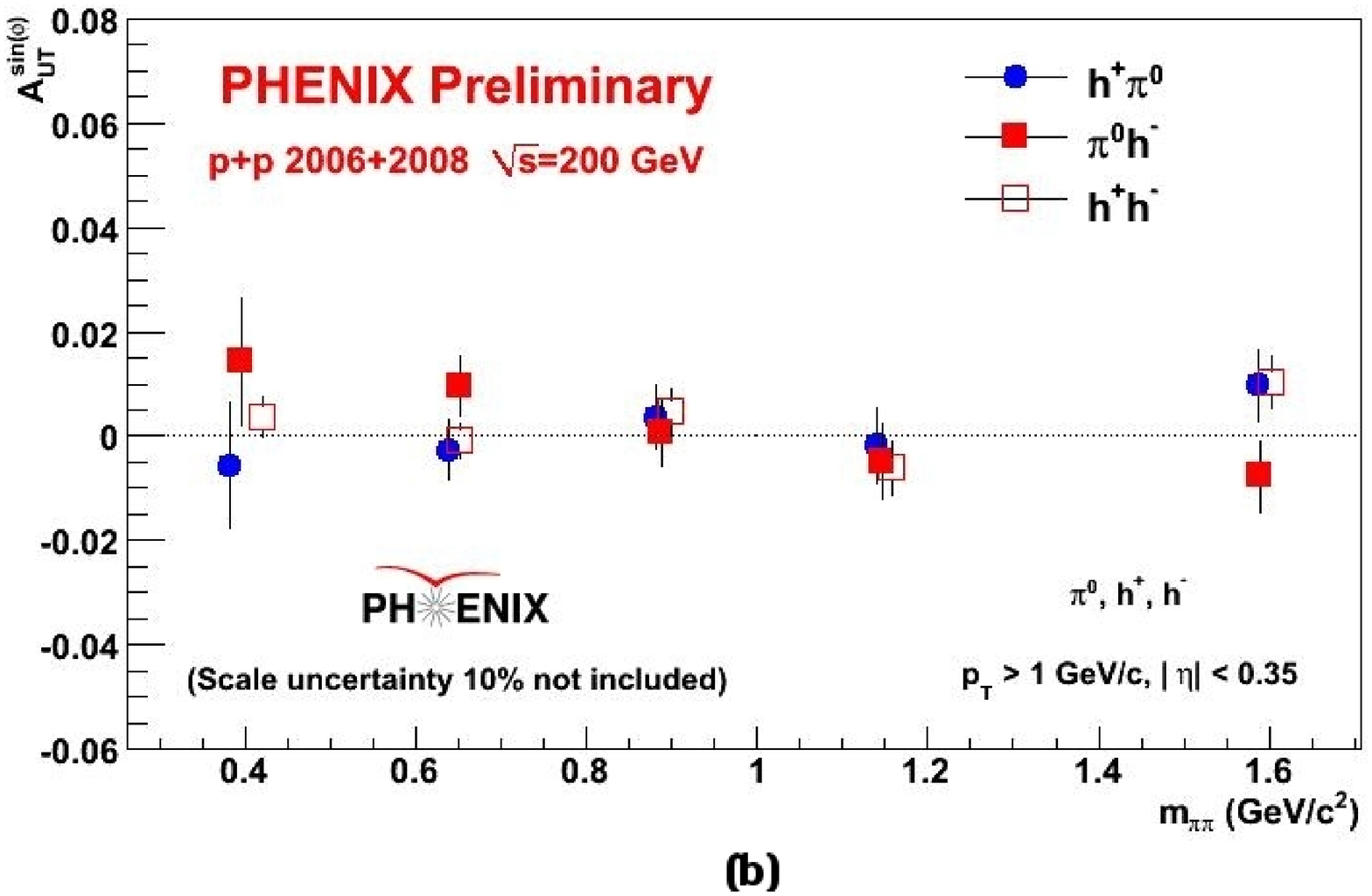, width=0.5\textwidth}
  \caption{(a) $A_N$ for back-to-back di-hadron correlation at mid-rapidity. (b) $A_N$ for di-hadron production within the same jet cone at mid-rapidity.}
\label{fig_Run6_dihadron_Sivers_IFF}
\end{figure}

In addition to measurements of single particle production, new channels in di-hadron
production have been explored. At PHENIX, the azimuthal correlation of two hadrons
produced in back-to-back jets has been measured as Figure \ref{fig_Run6_dihadron_Sivers_IFF}-(a) shows, which directly
probe the Sivers effect \cite{Bacchetta}. No significant effect has been measured with 2006 data.
This measurement can be expanded by correlating hadrons at forward rapidity and
with hadrons at midrapidity, which extend the sensitivity of the back-to-back di-
hadron Sivers measurement to higher and lower x.

The TSSA for di-hadron production within the same jet cone \cite{Boer} can be interpreted
as a product of the transversity distribution function and a spin-dependent di-hadron
fragmentation function, the so-called interference fragmentation function. As Figure \ref{fig_Run6_dihadron_Sivers_IFF}-(b)
shows, the measurement carried out with the combined dataset from 2006 and 2008
for hadron pairs at midrapidity showed that no significant asymmetry has been
observed with current statistical limits. In the near future this measurement will be
expanded into more forward kinematic region where valence quark will contribute
more in the pair production thus a larger asymmetry is expected.
\section{CONCLUSION AND OUTLOOK}
With detector upgrades for the PHENIX experiment and expected much improved
accelerator performance, RHIC-Spin program will provide a great opportunity for new
studies of polarized nucleon structure and the QCD dynamics. The silicon vertex
detectors in PHENIX will allow us to cleanly identify and separate leptons from heavy
quark and Drell-Yan decays. The future Drell-Yan $A_N$ measurement \cite{Kang2} is particularly
important at RHIC because the fundamental QCD predict that $A_N$ of Drell-Yan
should have an opposite sign of the $A_N$ observed in the polarized DIS experiment. Its
verification or disproval will be an important milestone in our understanding of QCD
spin physics. However, accumulating enough statistics is a challenging work for the
PHENIX experiment. The future high precision measurements of open heavy quark
and $J/\psi$ are expected as well.

\end{document}